\documentclass[aip,jcp,reprint,amsmath]{revtex4-1}
\usepackage{graphicx,dcolumn,bm,amssymb,paralist}
\usepackage{color}

\newcommand{\orderof}{\ensuremath{\mathcal{O}}}

\def\be{\begin{equation}}
\def\ee{\end{equation}}

\begin{document}
\title{Role of the potential landscape on the single-file diffusion through channels}
\author{S. D. Goldt and E. M. Terentjev}
\affiliation{Cavendish Laboratory, University of Cambridge, JJ Thomson Avenue, Cambridge CB3 0HE,
U.K.}

\begin{abstract}
  \noindent Transport of colloid particles through narrow channels is ubiquitous
  in cell biology as well as becoming increasingly important for
  microfluidic applications or targeted drug delivery. Membrane channels in
  cells are useful models for artificial designs because of their high
  efficiency, selectivity and robustness to external fluctuations. Here we model
  the passive channels that let cargo simply diffuse through them, affected by a
  potential profile along the way.  Passive transporters achieve high levels of
  efficiency and specificity from binding interactions with the cargo inside the
  channel. This however leads to a paradox: why should channels which are so
  narrow that they are blocked by their cargo evolve to have binding regions for
  their cargo if that will effectively block them?  Using Brownian dynamics
  simulations, we show that different potentials, notably symmetric, increase
  the flux through narrow passive channels -- and investigate how shape and
  depth of potentials influence the flux. We find that there exist optimal
  depths for certain potential shapes and that it is most efficient to apply a
  small force over an extended region of the channel. On the other hand, having
  several spatially discrete binding pockets will not alter the flux
  significantly. We also explore the role of many-particle effects arising from
  pairwise particle interactions with their neighbours and demonstrate that the
  relative changes in flux can be accounted for by the kinetics of
  the absorption reaction at the end of the channel.
\end{abstract}

\maketitle

\section{Introduction}

Transport of macromolecules through nano-sized pores and narrow protein channels
is essential for cell function \cite{alberts2008molecular} while also becoming
increasingly important in microfluidic applications \cite{shah2008,Pagliara2013}
or to understand drug delivery.\cite{Sugano2010} Channels in cell membranes are
remarkable for their high efficiency, selectivity and robustness with respect to
fluctuations of their environment \cite{hille2001ion} and come in two
flavours. Active transporters move their cargo by using cellular energy,
e.g. from hydrolysing adenosine triphosphate or by harvesting concentration
gradients of cell metabolites across the membrane. On the other hand, passive
transporters are driven by the growth of entropy of the system as they
translocate their specific cargo. Initially thought of as molecular sieves that
select via the pore size to let the `right' cargo simply diffuse through the
channel, it is now well established that passive transporters achieve high
levels of efficiency and specificity from binding interactions with the cargo
inside the channel. A well-characterised example is the bacterial channel
Maltoporin, where oligosaccharide transport is facilitated by an extended
binding region.\cite{Schirmer1995} Although many more examples of this
phenomenon have since been discovered using a plethora of methods (e.g. ex-situ
crystallographic studies,\cite{Kasianowicz2006} indirect measurements of ionic
currents \cite{Hilty2001} and molecular dynamics simulations,\cite{Jensen2002})
the exact details of the mechanisms of passive transporters are still poorly
understood.\cite{Kolomeisky2007}

Our work is motivated by a seeming paradox that arises when one considers the flux
through a narrow channel, such as Maltoporin, which prevents particles from
overtaking each other. Increasing the binding affinity between the channel
interior and the cargo will prolong the time each particle spends inside the
channel, hence reducing the flux and effectively blocking it. Why then would
channels evolve to have binding regions for the molecules they have evolved to
translocate? In this paper, we combine the results of Brownian dynamics
simulations with theoretical arguments to show how developing binding regions
inside a channel can indeed increase flux through narrow channels.
We model the binding regions using a variety of
potential-energy landscapes along the channel and investigate the dependence of
the flux through the channel on the shape and depth of these potentials.

We first consider a single freely diffusing particle to tune our Brownian
dynamics simulations in the setting where an exact analytical solution for the
transport exists, applying various tests to the simulation procedure to ensure
its proper reflection of the physical situation. We then investigate single-file
diffusion through a channel to analyse the dependence of particle flux on the
shape and depth of applied potentials. Finally, we demonstrate that we can
account for the relative changes in flux by considering the diffusion-limited
reaction kinetics of the absorption in a scheme \cite{Dorsaz2010} based on the
osmotic pressure along the channel alone.

To investigate the dynamics of colloid particles diffusing freely or in a
potential, we need to solve the Langevin or corresponding Fokker-Planck
equation. In a many-body problem like the one considered here, solving these
equations directly is completely unfeasible and we therefore turned to Brownian
dynamics (BD) simulations \cite{Chen2004} using the LAMMPS package.\cite{Plimpton1995}
 This allowed us to efficiently compute solutions of the
equations from an ensemble of simulated stochastic processes with the same
initial conditions.


\section{\label{sec:bp}Free Brownian particle}

The one-dimensional motion of a Brownian particle is described by the {Langevin equation}
$m\dot{v} + \alpha v = \zeta(t)$,
where $m$ is the particle mass, $v$ its velocity and $\alpha$ is the viscous
drag coefficient. We further assume white noise $\zeta(t)$ with
$\langle\zeta(t)\zeta(t')\rangle = \Gamma \delta(t -t')$, where $\Gamma = 2\alpha k_BT/m$ is the intensity of the stochastic force, satisfying the fluctuation-dissipation theorem. The general
solution for the root mean square displacement of the particle
is \cite{Uhlenbeck1930}:
\begin{equation}
    \label{eq:msd_general}
    \langle \Delta x^2\rangle = \frac{2k_BT\, m}{\alpha^2}\left(t/\tau_v - 1 + \exp(- t/\tau_v)\right)
\end{equation}
with the velocity relaxation time $\tau_v=m/\alpha$. In the overdamped (or
diffusive) regime, with $t \gg \tau_v$, we recover the Einstein result: $
\langle \Delta x^2\rangle = 2({k_BT}/{\alpha})t $, where one defines the
diffusion constant $D\equiv k_BT/\alpha$. On the other hand, in the {inertial
  regime} $t \ll \tau_v$, the displacement grows linearly with time: $ \langle
\Delta x^2\rangle = ({k_BT}/{m})t^2$.

\subsection*{\label{sec:bd}Brownian dynamics simulation}

To verify that LAMMPS yields
particle trajectories with the right statistical properties, we first simulated
freely diffusing spherical Brownian particles of different sizes at different
temperatures. The goal was to identify the crossover from the inertial to the
diffusive regime at different temperatures and to verify the
fluctuation-dissipation theorem.

Integration of the Langevin equation and the application of thermostat conditions
was done via the
\texttt{fix\_langevin} routine.\cite{Schneider1978} The free particles were simulated
in a box with periodic boundary conditions and were assigned
initial velocities drawn from a uniform distribution for the given
temperature. The viscous drag coefficient $\alpha$ was computed using Stoke's
law for a spherical particle at low Reynolds number: $\alpha = 6 \pi \eta R$
where $R = \sigma/2$ is the particle radius and $\eta$ is the fluid viscosity.

\begin{figure}
  \centering
  \includegraphics[width=\linewidth]{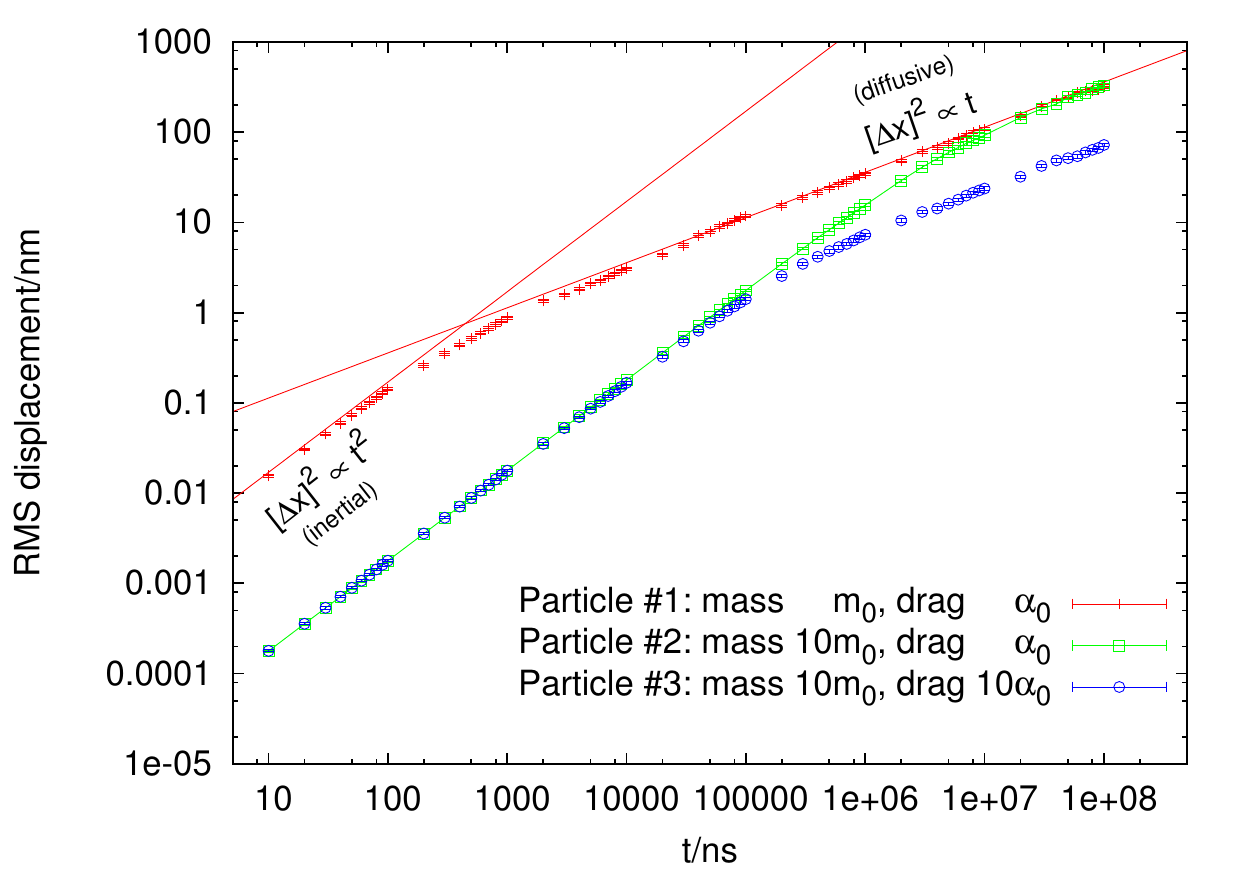}
  \caption{\label{fig:rmsd}\textbf{Inertial to diffusive crossover for the
      single particle.} Points are rms displacements computed from the
    simulated trajectories of 100 freely diffusing single particles
    for each set of parameters. Solid lines are the inertial and the
    diffusive limits of the Langevin
    equation solution \eqref{eq:msd_general}. Particle
    \#2 initially follows the trajectory of particle \#3, with has the same
    mass, before crossing over to the trajectory of particle \#1, which has the
    same drag coefficient. Errors were computed from the statistical distribution of particle
    displacements, but are to small to appear on this scale.}
\end{figure}

Figure \ref{fig:rmsd} shows the average root mean squared displacement of the
particles computed as the average of 100 simulated trajectories per
particle. Since there is no energy scale for the free particle, we used natural
units and in this case set the diameter of the particles to
$\sigma=1\mu\text{m}$ and their mass density to that of water, yielding a mass
of $m\sim4.2\cdot10^{-15}\text{kg}$. We found that the statistical relative errors for
the average trajectories were negligible for this number of simulations.

The crossover time was determined by inspection from the graphs. We can read off
a crossing-over time of $\sim 0.5\mu\text{s}$ for particle \#3 where the
inertial response has fully died down; this is on the same order as the
characteristic time scale $\tau_v = m/\alpha \sim 220 \text{ns}$.

We simulated particles of different sizes ($1-20\mu \text{m}$) at
different temperatures ($293-400\text{K}$). Here it may be necessary to account
for the fluid viscosity variation with temperature, and we used the empirical
formula for water \cite{Al-Shemmeri2012}:
\begin{equation}
  \eta(T) = 2.141\cdot10^{-5} \cdot 10^{247.8/(T-140)}.
\end{equation}
We have confirmed that the particle trajectories generated by LAMMPS had the
statistical properties expected from theory. We were also able to confirm that
the crossover time $\tau_v$ is practically independent on the heat bath
temperature, since the viscosity only depends weakly on temperature in the range
that we covered in our simulations.

\begin{figure}
  \centering
  \includegraphics[width=\linewidth]{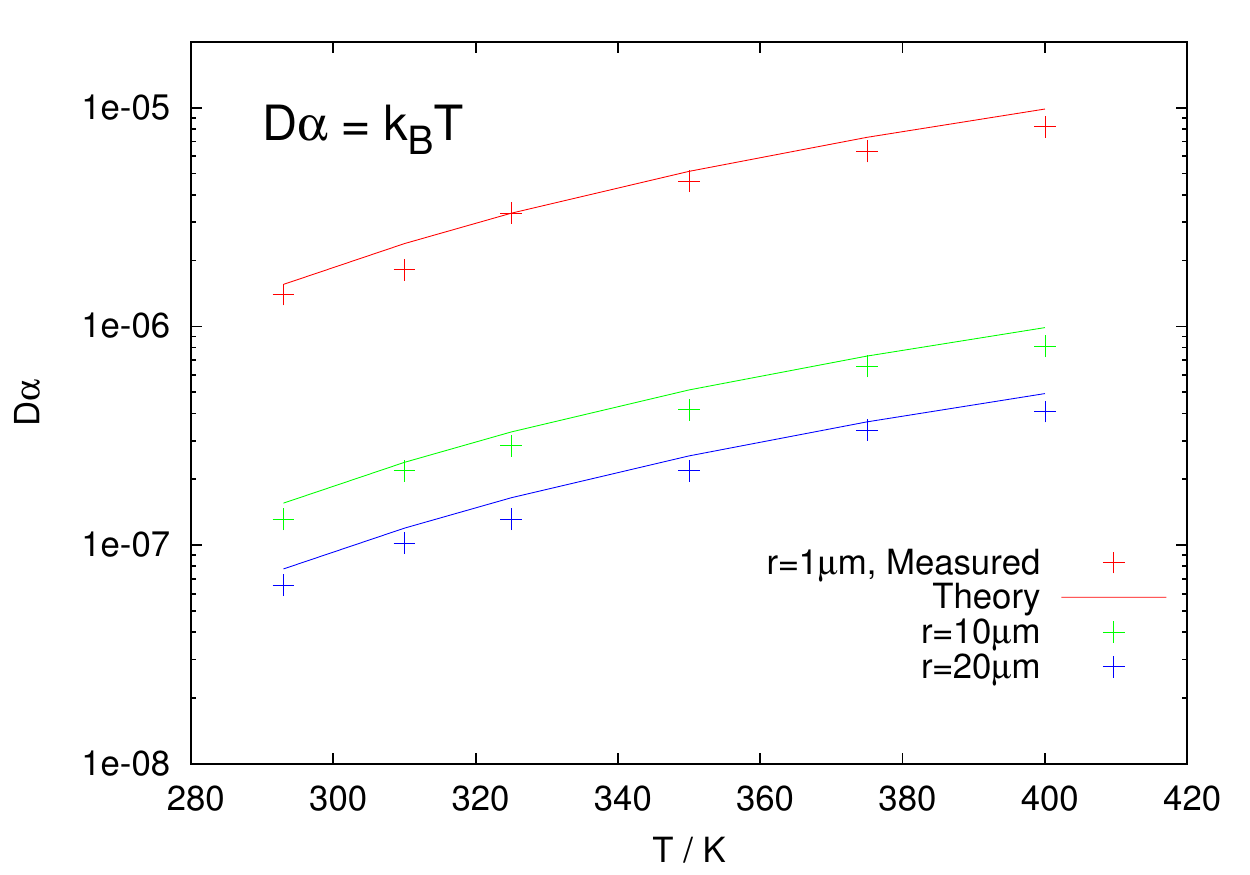}
  \caption{\label{fig:fd}\textbf{Testing the fluctuation-dissipation theorem.}
    Diffusion constants $D_{sim}$ were obtained from of average trajectories and
    are plotted as points, multiplied by the drag coefficient $\alpha=6\pi\eta
    \sigma$ which was given as a parameter to the \texttt{fix\_langevin} routine
    for simulations at different temperatures. The different colours correspond
    to the types of particles, with different radii as indicated on the
    plot. The corresponding Einstein relations $D\alpha = k_BT$ are plotted as
    solid lines.}
\end{figure}

To verify the fluctuation-dissipation theorem, which is used to derive the
Einstein relation $D=k_BT/\alpha$, we computed the diffusion constant from a
linear fit of the last three decades of each trajectory, i.e. for $t>10^5
\text{ns}$, obtained from 100 simulated particles using the
\emph{SciPy} \cite{Oliphant2007b} \texttt{curve\_fit} routine.  Figure
\ref{fig:fd} shows the product $D \alpha$ that was computed theoretically using
the Stokes relation (see above) as a function of temperature,
compared with the measured data.  The predicted trend is observed, with a very
small systematic offset that has been observed for a number of integration
schemes in Brownian dissipative dynamics.\cite{Vattulainen2002} Essentially,
this is an artefact that arises from the coarse-graining of the microscopic
properties of the fluid using a random force $\zeta$ while imposing overall
momentum conservation; this error is not significant for our purposes for two
reasons: the algorithm produces trajectories in almost perfect agreement with
the theory across a broad range of temperatures and furthermore, previous
studies have shown that the effects due to integrator artefacts are only
significant when the conservative forces of interest are comparable to the
thermal fluctuations \cite{Vattulainen2002} -- all the potentials we apply will
exceed energies of a few $k_BT$.

\section{\label{sec:free_channel}Free particle in a channel}

Having established the dynamics of Brownian particles and verified that LAMMPS
generates trajectories with the desired statistics, we now turn our attention
to the diffusion of free particles confined in a narrow channel.

We considered particles of diameter $\sigma = 2R$ freely diffusing through a
channel of radius $\sigma$. Note that from here on, we will use Lennard-Jones
(LJ) units which render all quantities dimensionless by assuming that particle
interactions follow the standard Lennard-Jones potential $V(r) = 4 \epsilon
[(\sigma/r)^{12} - (\sigma/r)^6]$ and setting the particle mass $m$, the
Boltzmann constant $k_B$ and $\epsilon$ and $\sigma$ as defined above equal to 1
\cite{ljunits}. All masses are to be understood as multiples of these
fundamental values, while other variables can be transformed to a dimensionless
form by a scaling with an appropriate combination of the above, e.g. for time:
$t_{\mathrm{LJ}} = t \cdot \sqrt{\epsilon m^{-1}\sigma^{-2}}$. For a full list
of conversion formulae see.\cite{ljunits} LJ units are widely used in
computational physics and offer the advantage of being able to treat systems of
different size and energy scales in one framework.

In our simulations, particles are modelled as spheres with a Lennard-Jones 12/6
type repulsion and no attractive interaction tail, that is, the LJ potential of
pair interaction is truncated at the point of its minimum, $r^* =
2^{1/6}\sigma$. The channel radius is too small to allow particles to overtake
each other, thus producing the \emph{single-file diffusion} and reflecting the
experimental fact that many metabolites will completely block their channels
during the transport due to their tight fit.\cite{Nestorovich2002}

\begin{figure}[htb]
  \centering
  \includegraphics[width=\linewidth]{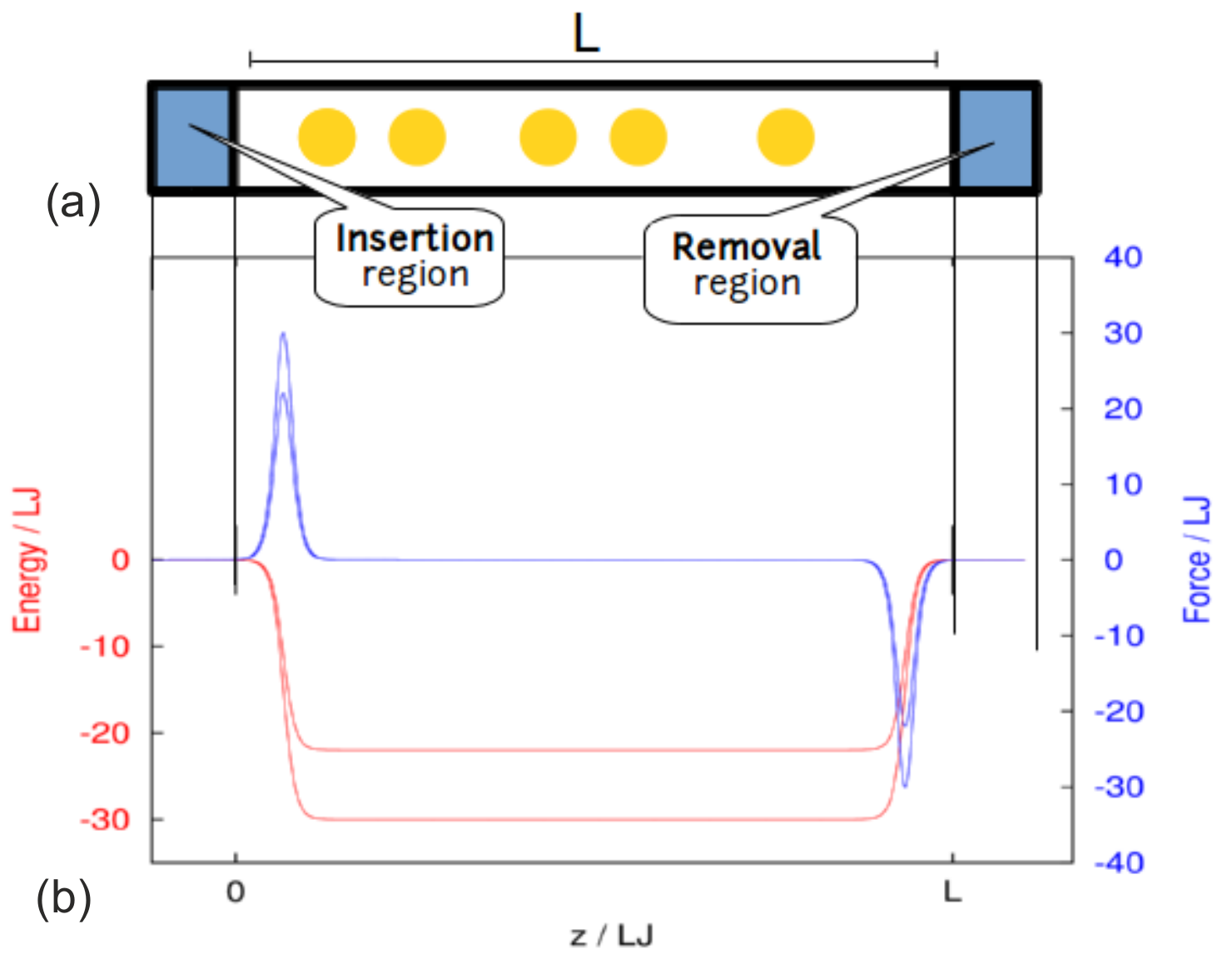}
  \caption{\label{fig:setup}\textbf{Channel geometry and an applied potential.} (a) The simulation box is split
    in three, with the `channel' of length $L$ in the middle aligned along the
    z-axis, containing spherical particles. Particles are inserted in
    the blue region to the left and removed from the simulation once they have
    crossed the channel and entered the blue removal region to the right. (b)
    A typical applied potential (red), in this case a uniform well, and the corresponding force exerted on a particle (blue), to scale.}
\end{figure}

Figure \ref{fig:setup}(a) shows a 2D-projection of the setup of the simulation
box (all simulations were carried out in full three dimensions). The `channel'
of length $L$ is the white area in the middle and is aligned along the
$z$-axis. Its walls interact with particles using the repulsive part of the
Lennard-Jones potential, same as described above, thereby `softly' preventing
contact. Particles are inserted in the insertion region to the left (blue) if
there is enough space. They then diffuse inside the cylinder. Once they have
crossed the channel and entered the removal region to the right of the channel
(blue), they are removed from the simulation. Underneath, in
Fig. \ref{fig:setup}(b) is a plot of two example potentials $V(z)$ and their
corresponding force landscapes, to scale. To investigate the first passage time
distribution, no potential was applied.

\subsection*{Distribution of first passage times}

\begin{figure}[htb]
  \centering
  \includegraphics[width=\linewidth]{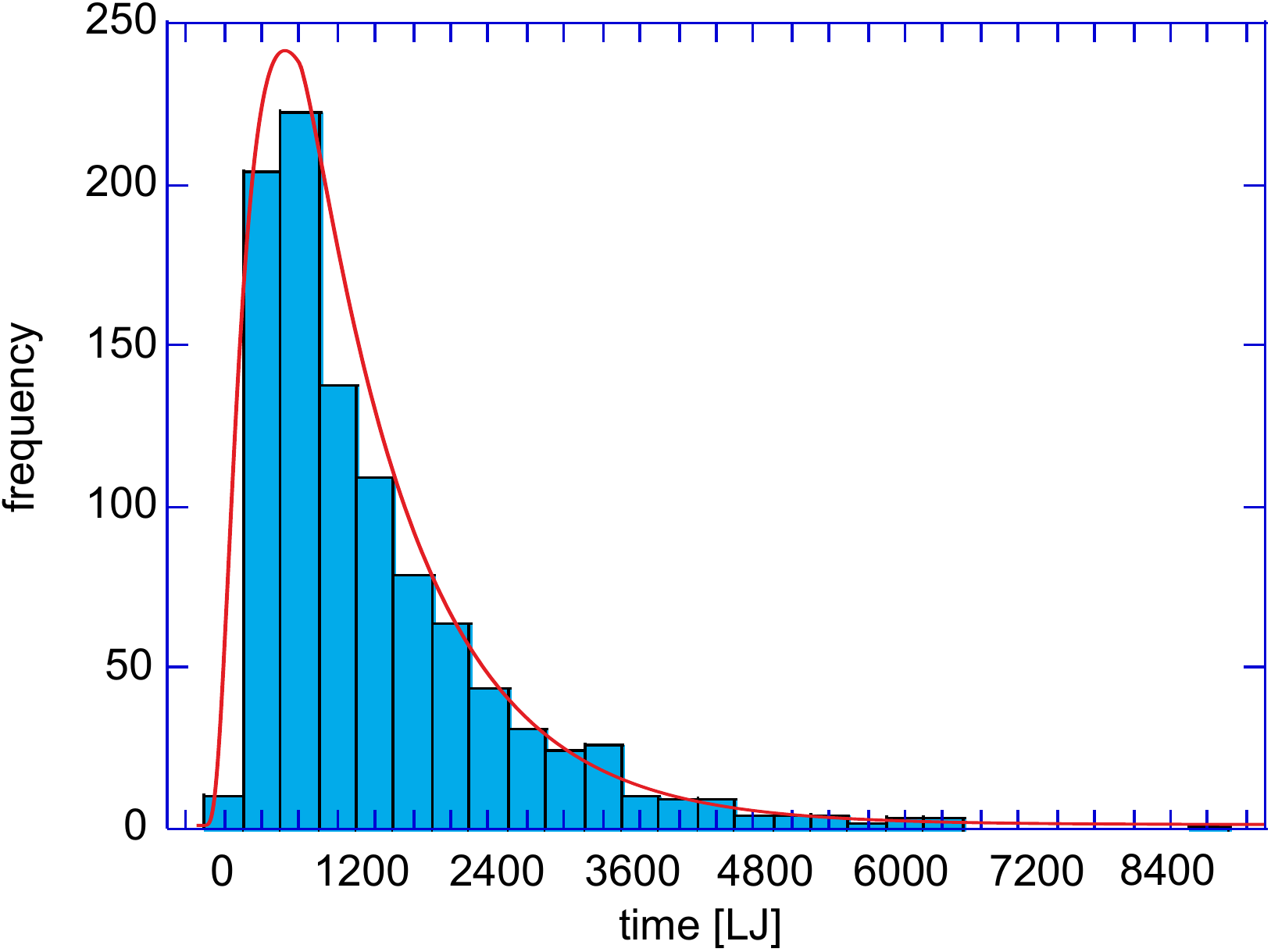}
  \caption{\label{fig:fp}\textbf{Distribution of first passage times
  for free diffusion.} The histogram was obtained from the first passage
    time of 500 single-particle simulations. The solid line is the theoretically calculated
    distribution $f(\tau)$, fitted using the diffusion
    constant $D = 0.284$ and a normalisation constant as the fitting parameters.}
\end{figure}

To check the physics of our channel setup, we looked at the distribution of
first passage times, $f(\tau)$, of particles freely diffusing through the
channel, i.e. {without} any applied potential. There is a classical result due
to Lord Kelvin, obtained by the methods of images \cite{Lappala2013} -- however,
it is only applicable when the particle is free to diffuse as far as necessary
to the left of $z=0$, while the first passage time is being tested by arriving
at $z=L$ to the right of its entry point, see Fig. \ref{fig:setup}(a). In our
case the passage is blocked to the left, so to find the probability for a
particle $p(z,t)$ one needs to solve the one-dimensional free diffusion equation
with the boundary conditions: reflective wall, $\nabla p=0$ at $z=0$, absorbing
wall, $p=0$ at $z=L$, and the initial condition for insertion:
$p(x,t=0)=\delta(z)$. The solution is:
\begin{equation}
  \label{eq:prob}
  p(z,t) \propto \sum_{n=0}^\infty \cos \left[ \frac{\pi z}{L} (n+1/2)\right] \exp \left( -\frac{\pi D^2(n+1/2)^2}{L^2} t \right),
\end{equation}
where the constant $C_0$ is determined by normalisation. The survival
probability for the particle to remain anywhere between 0 and $L$, having
started at $z=0$ is obtained by integration: $Q(t) = \int_0^L p(x,t) dx$. Given
the boundary conditions, $Q(t)$ does not depend on anything happening outside
the $(0-L)$ interval. Given the definition of the survival probability, the
fraction of particles equal to $-dQ(t)/ dt$ is absorbed between $t$ and
$t+dt$. This means that $f(t)=-dQ(t)/dt$ is actually the probability density of
the time $t$ that takes the particle to reach $z=L$ for the first time. This
distribution function is plotted in Fig.\ref{fig:fp}, and it gives
\emph{average} first passage time $\tau_\text{diff} \approx 4.92 L^2/\pi^2D$.

We measured the passage times in 500 simulations of a single particle diffusion
(to ensure no pair-interaction events could take place), where we inserted a
particle in the insertion area and allowed it to freely diffuse to the end of
the channel, the time for which we measured.  We fit the distribution of first
passage times to the resulting histogram of particle travel times in
Fig.\ref{fig:fp}, where only the diffusion constant $D$ and a normalisation act
as fitting parameters. The data are in good agreement with the model, and we
find a bare diffusion constant $D_0=0.28$.

Interestingly, we found that when we considered a similar experiment where we
inserted several particles into the channel one by one at a certain low
frequency, the distribution of first passage times severely deviated from the
free-diffusion result even at concentrations of just 2-3 particles inside a
channel of length $L=30$ at any one time. This shows that many-particle effects
caused by particle-particle interactions cannot realistically be ignored even at
the lowest of concentrations, a point to which we will return at the end of this
paper. In this particular case, the average first passage time was significantly
increased at these concentrations, suggesting a smaller diffusion constant or
higher effective resistance.

\section{\label{sec:applying_potentials}Potential along the channel}

Having established that our simulation setup produces physically meaningful
results, we now turn to the dependence of the flux through a passive channel on
the potential landscape inside it. We therefore made a series of experiments, in
each of which we simulated the insertion of 100 particles in the channel at a
rate $0.001\tau^{-1}=1/T_\text{in}$. Since this is now a genuine multi-particle
problem, simulation time increases accordingly. We therefore made use of the
parallel computing capacities of LAMMPS.

\begin{figure}[htb]
  \centering
  \includegraphics[width=\linewidth]{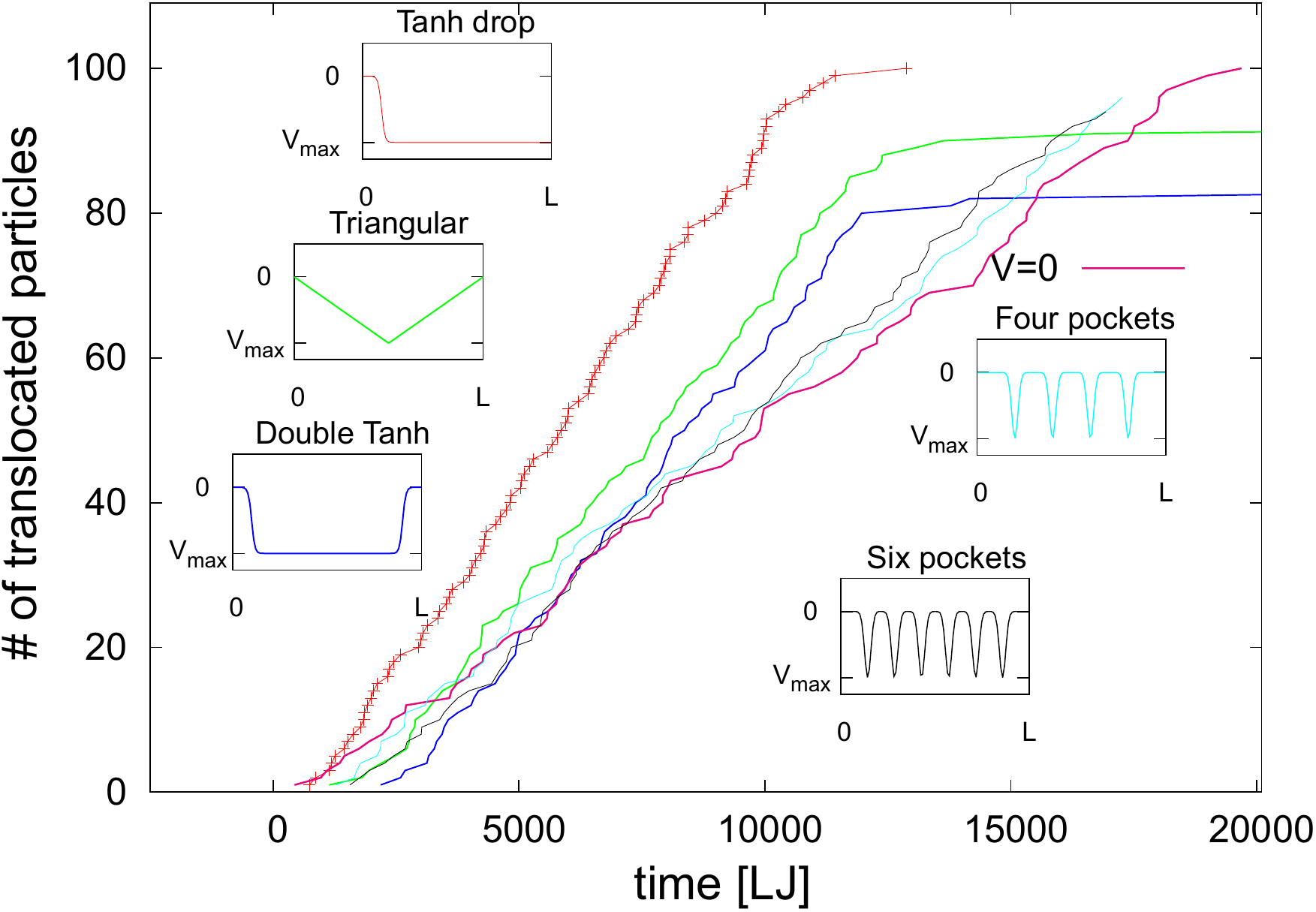}
  \caption{\label{fig:translocation}\textbf{Cumulated number of translocated
      particles for different potential shapes with $V_\text{min}=-20$.} The
    insets show the applied potential in the colour of the translocation curve
    that it produced, e.g. the `Double tanh' potential, plotted in red, was
    applied along the channel in a simulation that yielded the red translocation
    curve. Discrete binding pockets don't change the flux compared to the free
    ($V=0$) channel, plotted in pink. Symmetric potentials (blue, green) enhance
    the flux, but as we'd expect not as much as a simple potential drop at the
    beginning of the channel (red).}
\end{figure}

Particles were only inserted if there was enough space in the insertion region;
if a particle could not be inserted due to crowding at the channel entry, the
insertion was skipped and the next attempt was made after a time interval of
$T_{in}$. Inside the channel, one out of a number of different potential
shapes was applied with potential depth between $V_\text{min} \in [-5,-70]$ in LJ
units. The shapes of the potential are shown in the insets of
Fig. \ref{fig:translocation}. Note that potential `steps' are modelled using
$\tanh$ functions, hence the names `Double tanh' etc. Potential shapes include
`continuous' potentials, where the channel is modelled having a homogeneous
attractive interaction along its entire length (single/double tanh, triangular
potentials) or `discrete' potentials, where the channel provides a number of
discrete, spatially well-defined binding pockets, modelled as a Gaussian with a
depth $V_\text{min}$ and a standard deviation of $0.5\sigma$.

\begin{figure}[htb]
  \centering
  \includegraphics[width=\linewidth]{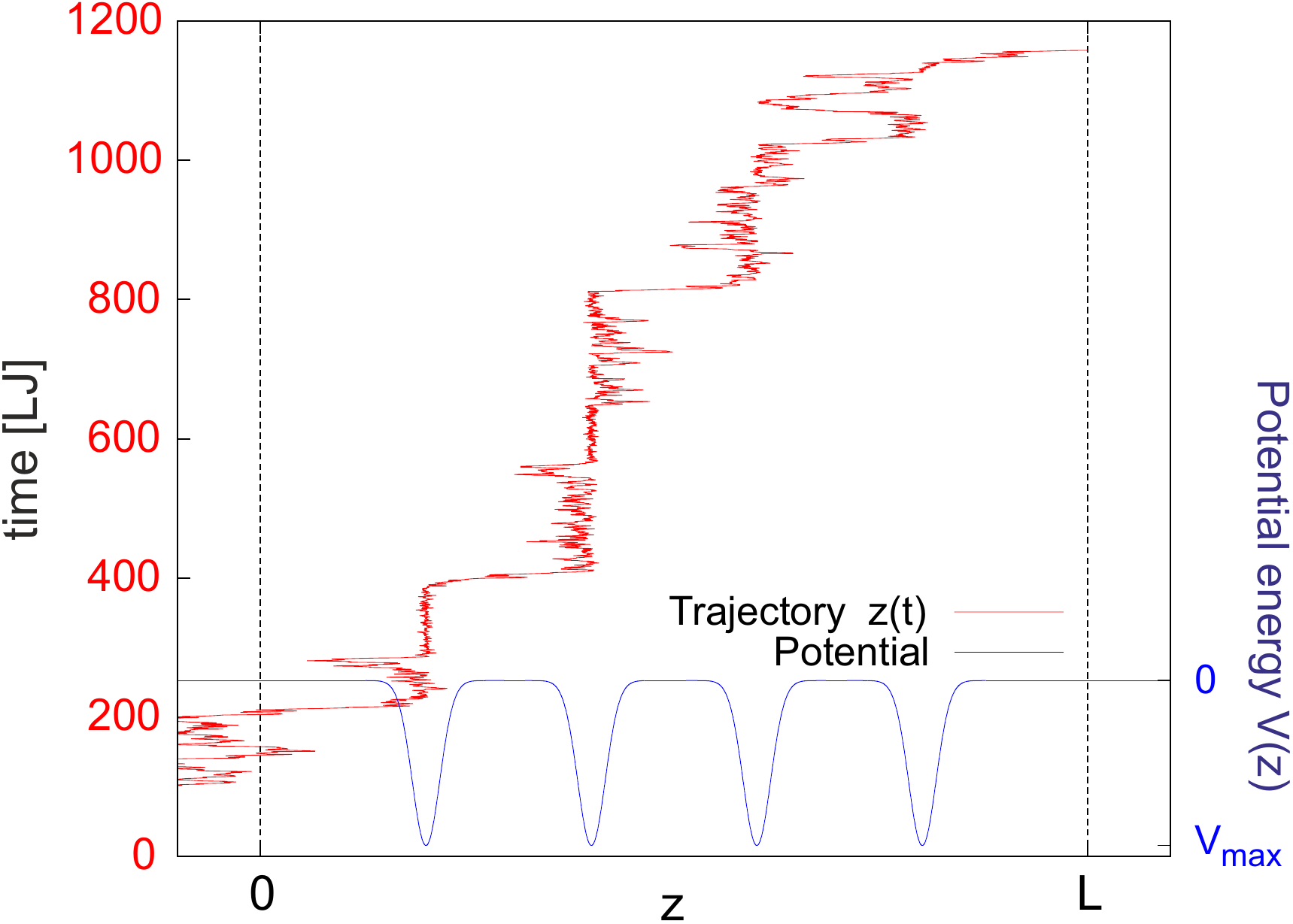}
  \caption{\label{fig:trapped}\textbf{Example trajectory of a single particle in
      a channel with four binding pockets.} Time is plotted on the y-axis and
    displacement along the channel in red on the x-axis. The potential inside
    the channel (blue) has four binding pockets, and the trajectory clearly
    displays trapping of the particle in every pocket, spending most time in the
    second pocket. The overall flux, however, is unaltered compared to a channel
    with no potential at all (see Figs. \ref{fig:translocation} and
    \ref{fig:flux_comparison}).}
\end{figure}

Despite the binding pockets being narrow, we can clearly see particle trapping
occurring by looking at individual particle trajectories such as the one shown
in Fig. \ref{fig:trapped}, where displacement of a single particle along the
channel is plotted on the x-axis in red, with time on the y-axis. The applied
potential, a series of four spatially discrete binding pockets, each modelled as
a Gaussians, is plotted in blue. We can clearly see that the particle is trapped
by every binding pocket, spending most time in the second pocket from the
left. However, the continuous insertion of particles to the left of the channel
leads to a consistent movement of the tagged particle to the right.

We measured the cumulative number of particles that have crossed the channel as
a function of time. An example of these measurements is shown in
Fig. \ref{fig:translocation}, where the cumulative number of translocated
particles is plotted as a function of time for different potential shapes, all
with the same $V_\text{min} =-20$. Insets show the exact form of the
applied potentials for each experiment.

We can already make a number of observations from
Fig. \ref{fig:translocation}. First of all, a simple potential drop at the
entrance of the channel leads to the expected acceleration and hence
significantly increased the flux (red). Symmetrical potentials produce an
increased flux compared to a channel with no potential ($V=0$) or different
numbers of discrete binding pockets, which surprisingly produce a flux on par
with the $V=0$ case. The number of translocated particles saturates at a value
lower than 100, the total number of particles inserted during the simulation,
because as particle insertion stops, exiting the channel becomes increasingly
harder for the remaining particles since the pressure inside the channel
decreases. The fact that the translocation number for a triangular potential
saturates at a higher value than for the double-tanh potential supports this
interpretation, since the double-tanh potential has a steeper wall at the end of
the channel which will effectively block the channel.

\begin{figure}[htb]
  \centering
  \includegraphics[width=\linewidth]{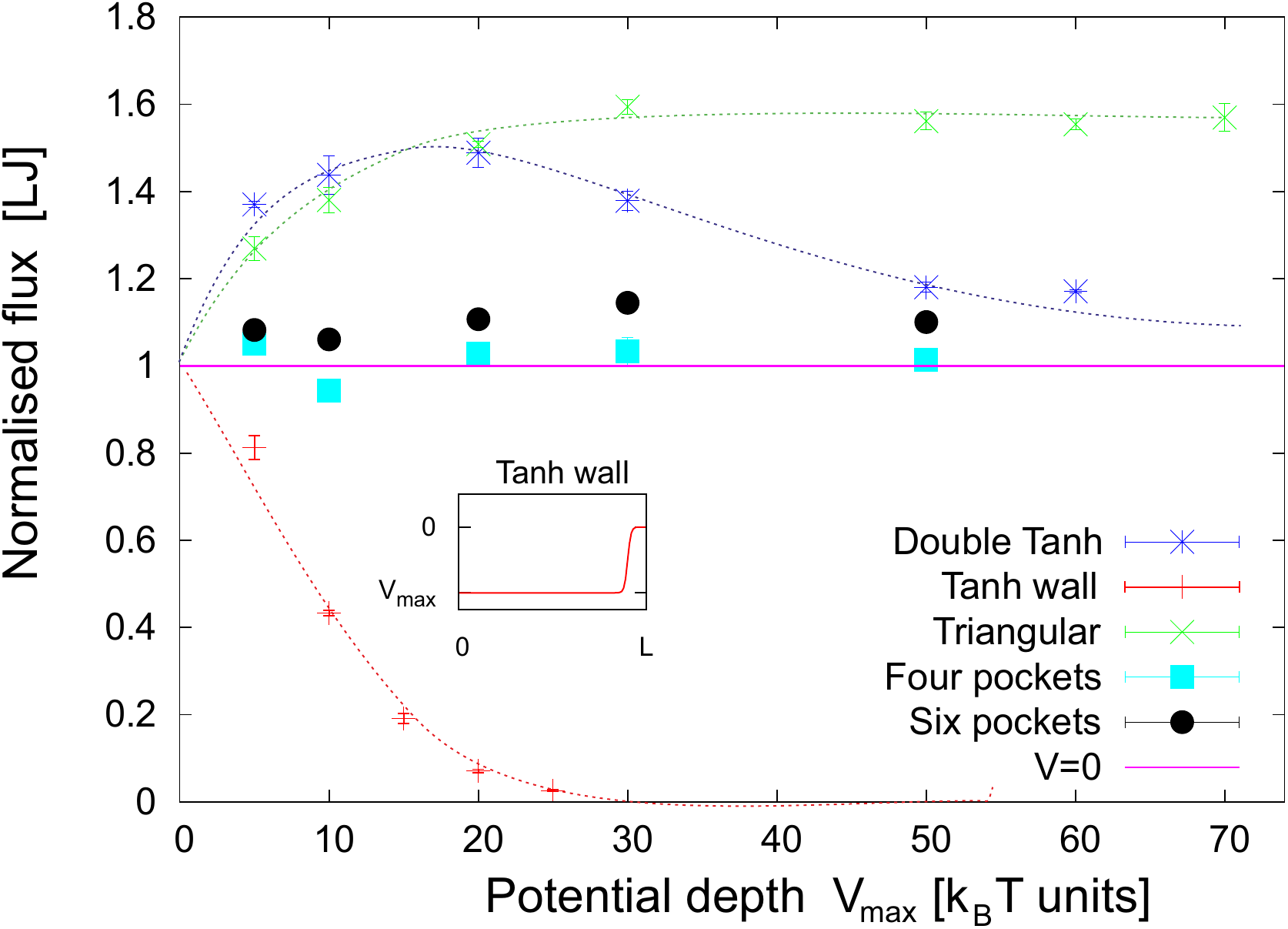}
  \caption{\label{fig:flux_comparison}\textbf{Normalised flux as a function of
  potential depth for different profiles.}
    Fluxes are defined as the slopes of a linear fit to the cumulative plot of
    the number of translocated particles as a function of time
    (e.g. Fig. \ref{fig:translocation}). Fluxes were measured from five
    experiments per potential shape/depth that simulated the
    translocation of 100 particles each and were normalised to the flux
    through a channel with $V=0$.
    The dashed lines for the three characteristic types of behaviour are to guide an eye. }
\end{figure}

Let us now discuss how the flux through the channel depends on the depth of
the potential for the different potential functions discussed so far. We
therefore define flux as the slope of a fit to the linear portion of the
translocation plots in Fig. \ref{fig:translocation}:
\begin{equation}
  \label{eq:flux}
  J = \left\langle \frac{\operatorname{d}\!N}{\operatorname{d}\!t} \right\rangle ,
\end{equation}
and repeat the analysis above for different potential depths. The average flux
was computed from five experiments for given potential shape and depth, that simulated
the translocation of 100 particles each. Simulating this number of particles
stabilised the linear fits and resulted in the small relative errors that are
plotted in Fig. \ref{fig:flux_comparison} and enable us to refine some of the
observations already made.

\noindent 1. \ Symmetric potentials increase the flux of single-file diffusion.
This is surprising at first sight, since the overall work done on the particle
is zero. However, the symmetry of the potentials is broken by the pressure that
the newly inserted particles exert on the particles near the end of the channel
at the potential wall.\cite{Lappala2013} It should be noted that this pressure
emerges purely from the free diffusion inside the channel and has significant
effects even at low colloid concentrations inside the channel. We are only
inserting particles if there is free space at the beginning of the channel, as
described above, so we are not actively pushing particles through the channel.

Furthermore, the flux through channels with symmetric potentials does not go
below the flux through a free channel even for deep potentials.

\noindent 2. \ The triangular potential profile outperforms the double-tanh
potential. This is due to the fact that in the overdamped limit, after an
impulse is exerted on a particle, it quickly relaxes back to normal
diffusion. Effectively, Newton's second law does not hold anymore and a small
force over a longer time, pushing the dense region forward at the entry half of
the channel, is more effective than a strong force over a short period of time.

\noindent 3. \ The increase in flux with symmetric potentials is not due to some
sort of Kramers-type barrier hoping. This is shown by the fact that the flux
through a channel with a $\tanh$ step at its end (`tanh wall') (see inset in
Fig. \ref{fig:flux_comparison}) goes to zero for $V_\text{max}=25$, where
double-tanh and triangular potentials still outperform the $V=0$ case.

\noindent 4. \ Narrow binding pockets do not alter the flux significantly, even
though it is clear from individual particle trajectories that particles do get
trapped in the binding pockets, as we see in Fig. \ref{fig:trapped}.

\begin{figure}
  \centering
  \includegraphics[width=0.85\linewidth]{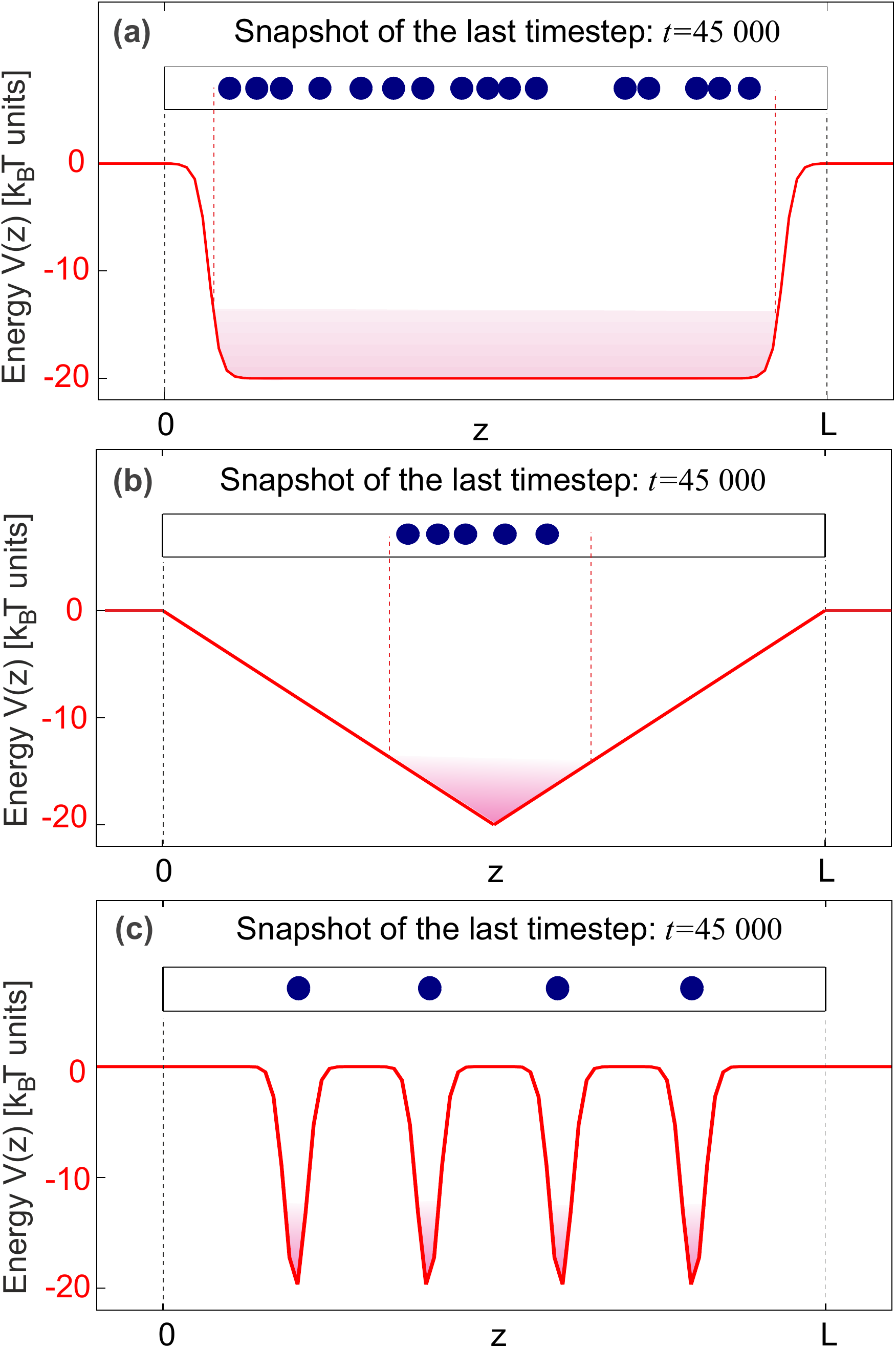}
  \caption{\label{fig:end}\textbf{Binding potentials keep particles in the
      channel.} The snapshots (a), (b) and (c) for different potential profiles
    give particle positions at the end of a simulation run, with channel and
    particle diameters to scale. This illustrates that sufficiently deep
    attractive potential will retain some particles, in regions shaded in the plots, when no additional influx
    from the left occurs (explaining the saturation plateau below 100 in
    Fig. \ref{fig:translocation}).}
\end{figure}

Figure \ref{fig:end} gives the snapshots of final simulation frames to
illustrate what is an `equilibrium' situation in each potential profile
$V(z)$. It shows that for a sufficiently deep attractive potential well,
particles are retained in such a well, while the particles facing weaker binding
forces escape and diffuse out of the channel. The final number of retained
particles explains why the plateaus of different curves in
Fig. \ref{fig:translocation} are below 100. These snapshots also help understand
why the flux increases with the depth of continuous potentials (double-tanh or
triangular). The process is analogous to the enzymatic action: although the
energy barrier at the end of channel is prohibitively high (as illustrated by
the complete vanishing of diffusive flux for the `tanh-wall' potential in
Fig. \ref{fig:flux_comparison}), when particles are confined at a high density
in front of such a wall -- they are forced to escape, pushed by the neighbours
from the left.

It is also interesting to observe that at a constant temperature of our heat
bath, when these potentials become excessively deep, the channel does get
blocked: this occurs at $V_\text{max} > 50-60$ for the double-tanh potential,
and has to be inferred to occur at a much greater depth for the triangular
potential, see Fig. \ref{fig:flux_comparison}.

\section{\label{sec:theory}Theoretical considerations}

We would like to gain a better insight into the data shown in
Fig. \ref{fig:flux_comparison} from a theoretical perspective. Since the basic
features of single-file diffusion have already been discussed extensively
elsewhere,\cite{Levitt1973,Kollmann2003,Burada2009} we will focus our discussion
here on the relative change in flux through a channel when we apply a potential.

It turns out that an effective way to pose this problem is to describe
translocation as a reaction $A+B\rightarrow B$ where the colloidal particles $A$ are
absorbed by a ``trap'' $B$, i.e. the channel exit, upon encounter. The problem
of finding the flux through the channel becomes the problem of computing the
rate $\kappa$ of this reaction in a crowded single-file environment with applied
potentials. For systems with spherical symmetry in the limit of infinitely
diluted reactants $A$, this is a classical problem of diffusion-controlled
reaction kinetics, which was solved exactly by Smoluchowski,\cite{Smoluchowski1917}
producing the rate $\kappa_s = 4\pi D_0 \rho_0$ where $\rho(r)$ is the density
profile of reactants $A$ around the trap reaching the value $\rho_0$ at infinity.

In general, the reaction dynamics is governed by the diffusive Fokker-Planck
equation:
\begin{equation}
  \label{eq:fp_rape}
  \frac{\partial \rho}{\partial t} = D_c \nabla \cdot \left(\nabla \rho - \frac{F}{k_BT} \rho \right) ,
\end{equation}
which takes the many-body effects into account through
the inhomogeneous density profile $\rho(r)$ along the channel, which
  generates
an osmotic pressure $\Pi (r)$ that acts as to spread the density profile via a
  force per particle $F = -1/\rho(r)\cdot \nabla \Pi(\rho)$.\cite{Zaccone2012}
These two parameters are non-linearly coupled via the collective diffusion
coefficient $D_c=D_0 \cdot \partial \Pi / \partial \rho$, which makes an exact
solution of this problem a formidable task even for numerics and makes us look
for reasonable approximations.\cite{Zaccone2013}

\begin{figure}
  \centering
  \includegraphics[width=0.85\linewidth]{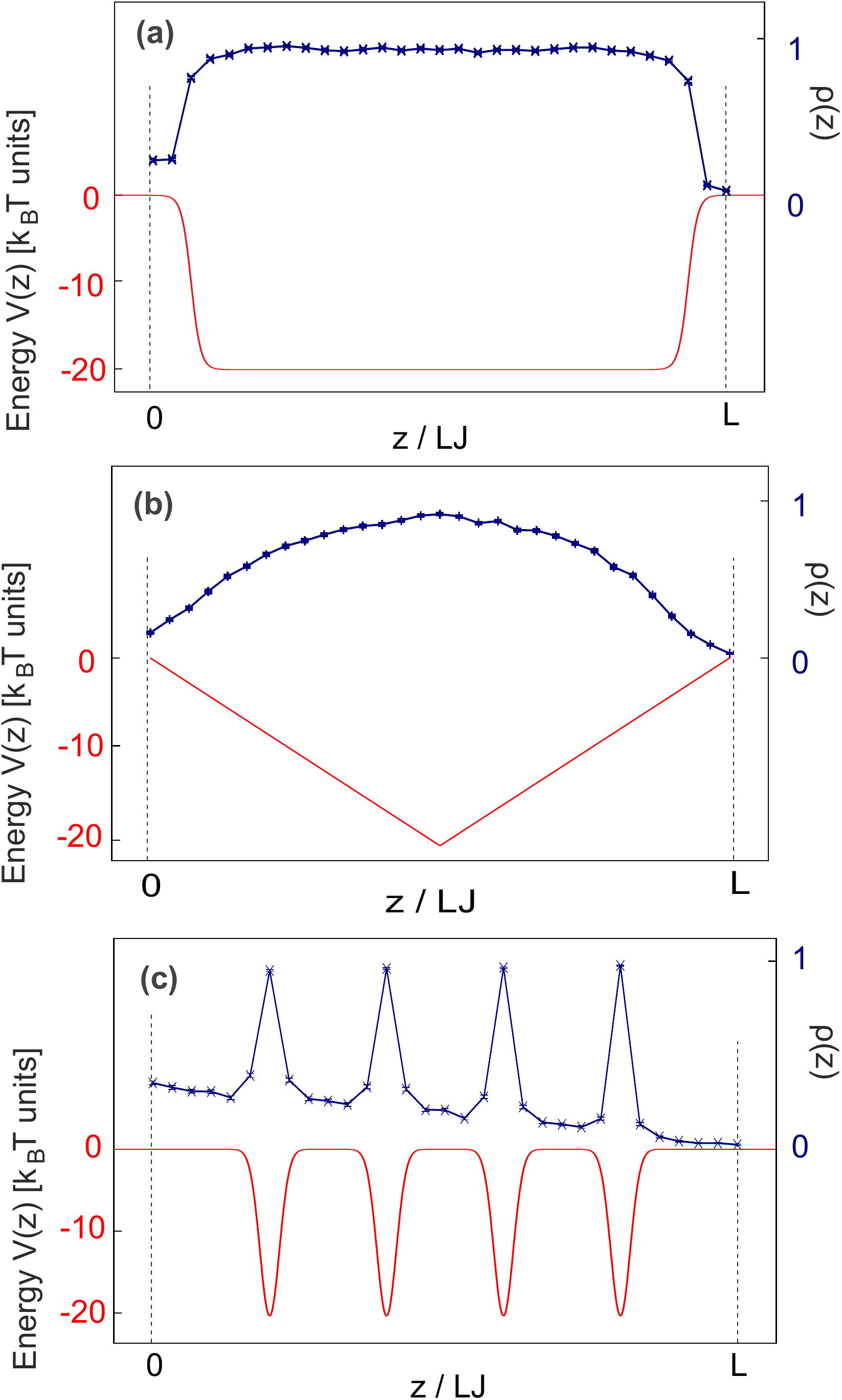}
  \caption{\label{fig:density}\textbf{Binding potentials keep steady-state
      density highly non-linear.} The snapshots (a), (b) and (c) for different
    potential profiles give the average density of particles in each case, in
    the steady-state transport regime (constant flux).}
\end{figure}

In our 1-dimensional case, we take $\rho (z)$ as the number of colloid particles
per unit length along the
channel at a given time (hence always $\rho < 1$ for narrow channels), and we
initially ignore the force due to the applied potential. With increasing $\rho$,
 it takes a given particle longer to reach the channel exit, but once
it is in the vicinity of the exit, its chance of reaching it and escaping increases
due to the added gradient of osmotic pressure. Using this model for the analysis of
their simulations, Dorsaz and co-workers \cite{Dorsaz2010} showed that the reaction rate can be
approximated well by the following expression:
\begin{equation}
  \label{eq:rate}
  \kappa \approx \kappa_s \frac{\beta \Pi(\rho_0)}{\rho_0} \cdot \exp\left(\frac{-\beta \Pi(\rho_\Delta)}{\rho_\Delta}\right)
\end{equation}
where $\beta=1/k_BT, \ \rho_0$ is the density at the beginning of the channel, and $\rho_\Delta = \rho(\Delta)$ is the density a characteristic `encounter distance' $\Delta$ from the channel end at which the density of colloid particles acquires structure due to interactions (in other words, where the ideal-gas linear relationship $\Pi = k_BT \rho$ stops being valid), see Fig. \ref{fig:density}. Equation \eqref{eq:rate} has since been derived
from first principles by Zaccone,\cite{Zaccone2013} who finds a prefactor of
$\beta (\operatorname{d}\!\Pi / \operatorname{d}\!\rho)_{\rho_0}$ instead
of $\beta \Pi(\rho_0)/\rho_0$, but notes that that the two prefactors
have the same dependence on $\rho_0$ which would indicate that the solution
is qualitatively correct.

We write the osmotic pressure $\beta\Pi$ as a virial
expansion in the density along the channel:
\begin{equation}
  \label{eq:osmotic_pressure}
  \beta\Pi = \rho + B_2\rho^{2} + B_3\rho^{3} + \orderof\left(\rho^{4}\right)
\end{equation}
and compute $\rho$ from 1000 randomly selected snapshots of the simulations after the
flux has equilibrated to its steady-state value. The virial coefficients $B_2$ and
$B_3$ account for two- and three-body interactions between the particles in the
channel which captures the essential dynamics since in the effectively 1-D
system of the channel, the motion of a particle is dependent on the particle in
front and the particle behind it.\cite{Lappala2013} $B_2$ and $B_3$ were
computed for a Lennard-Jones $12/6$ potential with $\epsilon=1, \sigma=1$ that
was used to model particle-particle interactions as described in section
\ref{sec:free_channel}. While $B_2 =2 \pi \int_0^\infty r^2 [1-e^{-\beta
  V(r)}]\operatorname{d}\!r$; computation of $B_3$ is more involved, but values
are available.\cite{Bird1950}

\begin{figure}
  \centering
  \includegraphics[width=\linewidth]{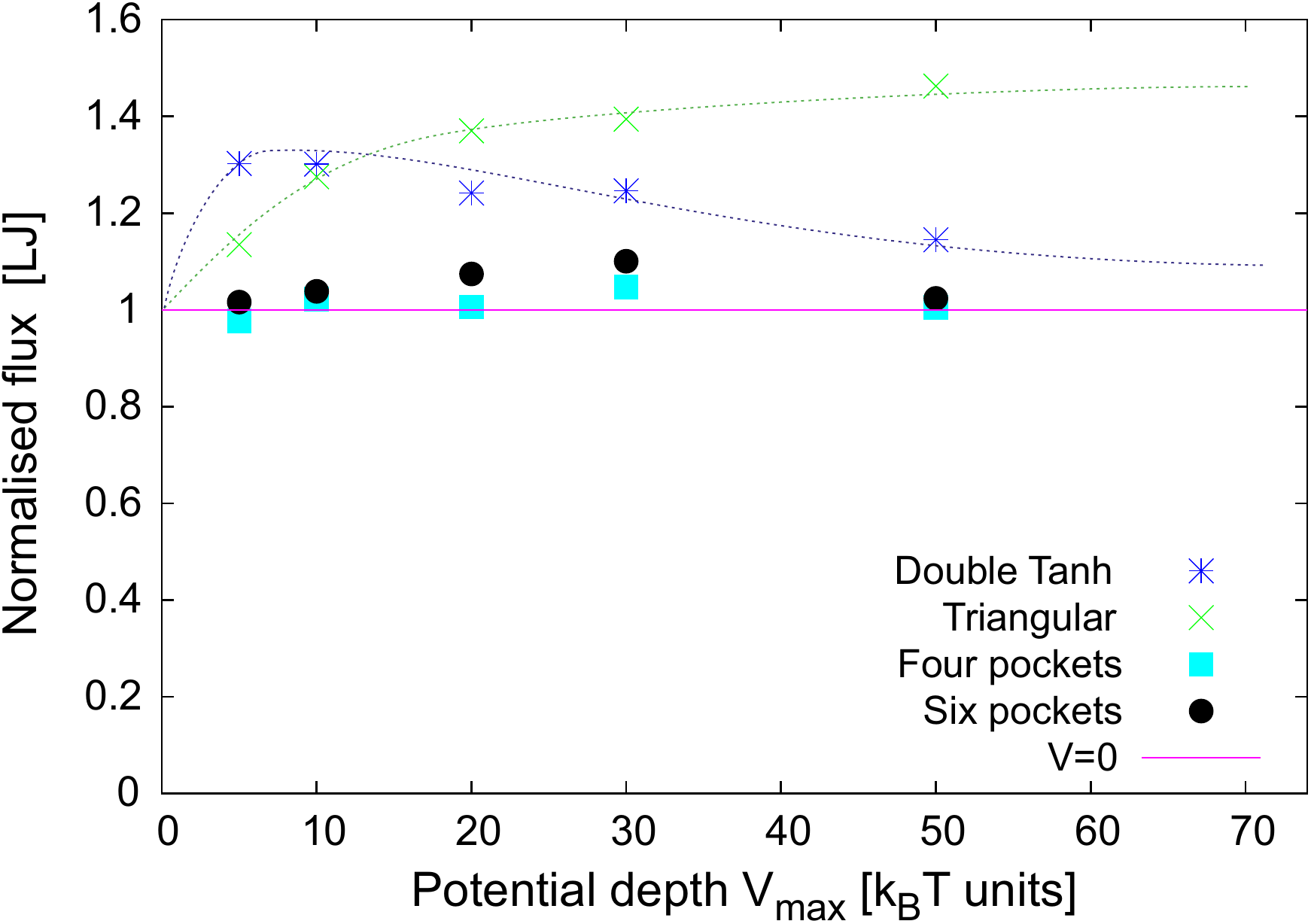}
  \caption{\label{fig:rate}\textbf{Reaction rates \eqref{eq:rate} for an
      absorption reaction $A+B\rightarrow B$ correctly predict the flux through
      a channel with applied potential.} The trends seen in the simulations
    (Fig. \ref{fig:flux_comparison}) are correctly predicted and numerical
    agreement is also good, although there is a systematic offset of $\sim 0.1$. Since
    the form of the potentials does not enter the model at any point, we conclude
    that all the information is encoded in the equilibrium density distribution,
    which we sample at only two discrete points.}
\end{figure}

A plot of the reaction rates computed from \eqref{eq:rate} is shown in
Fig. \ref{fig:rate}. All values are normalised with respect to the reaction rate
computed for no potential, $\kappa_{V=0} = 1$. It is clear from the graph that
these rates correctly predict the trends seen in the flux from the
simulations (Fig. \ref{fig:flux_comparison}): there is no significant flux
change with discrete pockets but a considerable increase with continuous
potentials; the double tanh potential performs best at small potential
depths while the triangular potential trumps at higher values of $V_{max}$. The
numerical range of the relative changes is good although it is systemically
low by $~0.1$. This is a remarkable agreement given that at no point we explicitly
introduced the form of the potentials and evaluate $\rho$ only at two discrete
points, \emph{i.e.}  the beginning of the channel and very close to its
exit. This shows that all the information about many-particle
effects and the channel translocation with an applied potential is encoded
in the steady-state density
distribution, which in turn is controlled by the two virial coefficients $B_2$ and $B_3$
(or the two values $\rho_0$ and $\rho_\Delta$ sampled near the beginning and near the end of the channel).

\section{Conclusion}

Our result that continuous, symmetrical potentials increase the flux
significantly confirms earlier speculation that membrane channels in cells are
most likely to provide a ``molecular slide'' \cite{Schirmer1995} by organising
discrete binding sites in succession, since having them isolated one after the
other would provide little to no increase in flux as shown. Furthermore, our
results can offer guidance for the design of artificial channels in microfluidic
applications, where improving flux is often important and clogging can be a
problem \cite{Lappala2013}.

Any theoretical description of particle translocation has to account for both
the applied potential and the crowding inside the channel. We have shown that it
is possible to account for the relative changes in flux by considering the
kinetics of the ``absorption reaction'' of particles exiting the channel, thus
mapping the many-body problem to a two-body-interaction where crowding is
modelled by the osmotic pressure inside the channel, without knowledge of the
applied potential. However, this is more of an explanation \emph{a posteriori}
since it requires knowledge of the density profile along the
channel. Further theoretical work will therefore have to focus on the
development of methods to calculate these distributions not just for periodic
boundary conditions\cite{Kollmann2003}, but for more realistic geometries and
boundary conditions in the presence of potentials in an attempt to predict
particle flux without resorting to simulations.

\section*{Acknowledgements}
We are grateful to Anna Lappala for many stimulating discussions.  Simulations
were funded by the Cavendish Laboratory teaching committee and performed using the Darwin Supercomputer of the University of Cambridge
High Performance Computing Service
(\href{http://www.hpc.cam.ac.uk/}{http://www.hpc.cam.ac.uk/}), provided by Dell
Inc. using Strategic Research Infrastructure Funding from the Higher Education
Funding Council for England.


\end{document}